
\normalbaselineskip=20pt
\baselineskip=20pt
\magnification=1200
\hsize 16.0true cm
\vsize 22.0true cm

\def\lsim{\mathrel{\rlap{\lower4pt\hbox{\hskip1pt$\sim$}}
    \raise1pt\hbox{$<$}}}         
\def\gsim{\mathrel{\rlap{\lower4pt\hbox{\hskip1pt$\sim$}}
    \raise1pt\hbox{$>$}}}         

\def\ut#1{$\underline{\smash{\vphantom{y}\hbox{#1}}}$}
\def\overleftrightarrow#1{\vbox{\ialign{##\crcr
    $\leftrightarrow$\crcr
    \noalign{\kern 1pt\nointerlineskip}
    $\hfil\displaystyle{#1}\hfil$\crcr}}}
\long\def\caption#1#2{{\setbox1=\hbox{#1\quad}\hbox{\copy1%
\vtop{\advance\hsize by -\wd1 \noindent #2}}}}

\centerline{\bf A Proposed Test of Charge Symmetry in $\Sigma$ Decay}
\vskip 24pt
\centerline{E.M. Henley}
\centerline{\it Department of Physics, FM-15 and Institute for Nuclear Theory,
HN-12}
\centerline{\it University of Washington, Seattle, Washington 98195}
\vskip 6pt
\centerline{and}
\vskip 6pt
\centerline{G.A. Miller}
\centerline{\it Department of Physics, FM-15}
\centerline{\it University of Washington, Seattle, Washington 98195}
\vskip 24pt
\centerline{Abstract}
The semi-leptonic decays of $\Sigma^\pm$ offer a vehicle for observing charge
symmetry-breaking.  The effect is expected to be about 6\%,
enhanced due to the
replacement of two $u$ quarks by $d$ quarks.  We propose that present
experimental data be improved to search for this effect.
\vfill\eject
In the recent 1st International Symposium on {\it Symmetries in
Subatomic Physics} Thomas$^1$ discussed enhanced charge symmetry-breaking
effects in
the difference between the valence distribution of down quarks in the
proton and up quarks in the neutron. The asymmetry is about 5-10 \% for
large values of Bjorken x,
larger about than expected
 because
there are two spectator u quarks in the proton and two spectator d quarks
in the neutron.

There is a simpler process
where the same enhancement occurs.  This is a comparison of the semi-leptonic
decay rates $\Sigma^+\to \Lambda^0 e^+\nu_e$ to $\Sigma^-\to\Lambda^0
e^-\bar\nu_e$.  Since the $\Sigma^+$ is $uu$s and the $\Sigma^-$ is $dd$s,
the rate comparison tests charge symmetry in a situation  where two up quarks
are replaced by two down ones.  

As shown by Frampton and Tung$^3$ and others$^4$, the non-leptonic decay rate
can be written as
$$\Gamma = {(G_F\sin\theta_c\;f_{\Sigma\Lambda})^2\over 2^615\pi^4(1+\delta)^3}
\;M^5_- (3|f_1|^2+5|g_1|^2)\;,\eqno(1)$$
where $G_F$ is the Fermi coupling constant, $\theta_c$ is the Cabibbo angle,
and $f_{\Sigma\Lambda}$ is an f-type SU(3) coefficient.  Also we have defined
$$M_- = M_\Sigma - M_\Lambda,\quad M_+ = M_\Sigma + M_\Lambda,\quad
\delta = M_-/M_+,\eqno(2)$$
$f_1$ is the matrix element of the weak vector current
 which we
assume to be a dipole  form factor $f_1=M^4_V(q^2+M^2_V)^{-2}$,
with $M_V\approx 0.84$ GeV, and $q$ the
4-momentum transfer in the decay. In Eq~.(1)
$g_1$ is the matrix element of the axial vector operator which we take
as a number $C_A$ times a form factor
, $g_1=M^4_A(q^2 + M^2_A)^{-2}$, where $M_A\approx 1.05 GeV$.
For $\Sigma^\pm$ decay $C_A$ is the matrix element of
$\bar {q} \gamma_\mu \gamma_5 \tau_\pm q$. If charge symmetry holds,
$$|\Sigma^+>=P_{cs} |\Sigma^->,\eqno(3)$$ where the
the charge symmetry operator$^2$
$P_{cs} = e^{i\pi T_2}$. This operator
converts $u$ quarks into $d$ quarks and vice versa: $P_{cs}|u>
= -|d>$, $P_{cs}|d> = |u>$. Charge symmetry leads to
the result that $C_A(\Sigma^+)=-C_A(\Sigma^-)$.

In this limit,
we obtain
$$R\equiv {\Gamma(\Sigma^-)\over\Gamma(\Sigma^+)} = {[M_-(\Sigma^-)]^5\over
[M_-(\Sigma^+)]^5}\;{[1+\delta(\Sigma^+)]^3\over [1+\delta(\Sigma^-)]^3}
\eqno(4)$$
to an accuracy of better than 0.1\%.
The deviation from  unity in Eq.~4
arises only from the difference in phase space factors
caused by the  mass difference between the $\Sigma^+$ and $\Sigma^-$.
The difference of the form factors $f_1$ and $g_1$
for $\Sigma^+$ and $\Sigma^-$ decay is negligible.  The use of Eq.~(4) and
known masses of $\Sigma^\pm$, $\Lambda^0$,$^5$ give
$$R=1.665\pm 0.009.\eqno(5)$$
At present, the decay rates are given by$^5$
$\Gamma(\Sigma^+) = (2.5\pm 0.6)\times 10^5 s^{-1}$,
$\Gamma (\Sigma^-) = (3.87\pm 0.18)\times 10^5 s^{-1}$. This gives
$$R^{expt} =1.6\pm 0.4.\eqno(6)$$

The present branching ratio of $\Sigma^-$ is measured to about 5\%, but that
of $\Sigma^+$ only to 25\%.
These errors are much larger than typical charge symmetry breaking effects,
which are typically of the order of 1-3\%.

It is worthwhile to provide estimates of the size of the charge symmetry
breaking effects of $C_A$.  The simplest effect to consider is that of
$|\Lambda^0>-|\Sigma^0>$ mixing$^6$,
The physical $|\Lambda>$ and $|\Sigma^0>$ are thought to be mixtures
of pure isospin states
$$
\eqalignno{|\Lambda^0> &= |I=0>+\alpha|I=1>, \cr
|\Sigma^0> &= -\alpha |I=0> +|I=1>, &(7)\cr}
$$
 with $\alpha\approx .013$ as estimated in Ref.~6.
The quark-model origin of this effect is the charge symmetry breaking
mass difference between up and
down  quarks which enters in the one-gluon exchange interaction.
We estimate the matrix elements of the vector $\hat V $ and axial-vector
$\hat A$ operators
between the $|\Sigma^\pm>$ and the physical $\Lambda>$, using SU(6)
wave functions.  The matrix element of
$\hat V$ between the
$|I=0>$ and $|I=1>$ pure isospin states vanishes, so the effects of
mixing in Eq.~(1) are of second order in $\alpha$ in the ratio $R$
and ignorable.
A simple evaluation of the matrix element of $\hat A$ reveals that
$$<\Lambda| \hat {A} |\Sigma^\pm>=-(\pm {2\over 3} \sqrt{3} -{4\over 3}\alpha
),
\eqno(8)
$$
so the ratio of the square of the matrix elements is
$$({g^+\over g^-})^2\approx 1 -{8\over 3}\sqrt{3}\alpha\approx 6\%.
\eqno(9)$$
This is a relatively large effect.
Other
charge symmetry breaking effects occur in the wave
functions, but these are much smaller.
The $\Sigma^+$ decay involves a u in the
$\Sigma^+$ changing into a d in the $\Lambda$, while $\Sigma^-$ decay
involves a d in the $\Sigma^-$ changing into a u in the $\Lambda$.
In the simplest bag or constituent quark model the u and d wave functions
of the $\Sigma$ and $\Lambda$ would be the same, but the quark mass
difference would cause the u
wave function to differ from that of the d.
However, that effect would give no
charge asymmetry here because there is a single u and a single d wave
function in both matrix elements. The charge asymmetry occuring via the effects
of spatial wave function is therefore
a subtle effect. In the
MIT bag model$^7$,
 a   mass difference $\delta m$
between the up and down quarks is required to explain
the mass difference between the neutron and proton.
Using $\delta m=4.3$  MeV$^8$ leads to the result
that 4 MeV ($\equiv \Delta M$)  of the 8.3 MeV
difference ($M_{\Sigma^+}> M_{\Sigma^-}$) must be supplied
by
electromagnetic and one gluon exchange effects.
The use of the bag model
stability equations gives
$\Delta R/{\bar R }= {4 \Delta M\over 3 \bar M}\approx 1/900$,
This is  barely perceptible, and therefore very sensitive to
a host of corrections. In any case, it is small and can therefore be
neglected.

We also make an estimate using the non-relativistic quark model.
The variational principle, with harmonic oscillator trial wave
functions$^9$, is used
 to simplify this first calculation. The result is that
$\Delta R/{\bar R }= {3 \Delta M\over 4 \bar M}$ for oscillator confinement
and $\Delta R/{\bar R }= {3 \Delta M\over 2 \bar M}$. Here $\bar R$ and
$\Delta R$
refer to the radius parameter of the harmonic oscillator wave functions.
Computing the relevant overlaps gives a difference of 0.6~\% (oscillator)
or 1.2~\%
(linear) for the square of
the matrix elements. There is extreme model dependence, but
the effect is much smaller than the 6\% expected from baryon
mixing.

We would like to urge increased analysis of
existing data or of new data to search for charge symmetry-breaking at the
1\%
level.  Since this symmetry tests the effects of isospin mixing in the
baryon  wavefunction,
it is of particular
interest.  No such asymmetry has yet been observed for a hadron.
\vskip 12pt
This work was supported in part by the U.S. Department of Energy.
A recent volume of Phys. Lett. B, containing Ref.~10,
 arrived in our library as we were about to
submit this for publication. Ref.~10 concerns the same topic and obtains
the same result for $\Sigma^\pm$ semi-leptonic decays.
\vskip 12pt
{\bf References}
\vskip 12pt

\item{1.} A. Thomas, at the 1st International Symposium on {\it Symmetries in
Subatomic Physics} to be published, ed. by W.-H P. Hwang;
E.N. Rodionov, A. W. Thomas and J.T. Londergan, ``Charge Asymmetry of
Parton Distributions", IU/NTC 93-19,ADP-93-222/T139

\item{2.} G.A. Miller, B.M.K. Nefkens, and I. Slaus,
Phys. Rept. 194, 1  (1990)

\item{3.} P.H. Frampton and W-K. Tung, Phys. Rev. D\ut{3}, 3 (1971).

\item{4.} See e.g. M. Bourquin {\it et al.}, Z. Phys. C{\ut{12}, 318 (1982).

\item{5.} Particle Data Group, Phys. Rev. D\ut{45}, S1 (1992).

\item{6.} R.H. Dalitz and F. Von Hippel, Phys. Lett. 10, 153 (1964).

\item{7.} T. De Grande, , R.L. Jaffe, K. Johnson, and J. Kiskis,
Phys. Rev. D12, 2060 (1975).

\item{8.} R. P. Bickerstaff and A.W. Thomas, Phys. Rev. D \ut{25}, 1869
(1982).

\item{9.} N. Isgur, p. 45, in ``Nucleon Resonances and Nucleon Structure",
ed. by G.A. Miller, World Scientific, Singapore , 1992.

\item{10.} G. Karl, Phys. Lett. B 328,149 (1994).
\bye